# The Convergence of Digital-Libraries and the Peer-Review Process

Marko A. Rodriguez[*§†], Johan Bollen[†], Herbert Van de Sompel[†]

[*]Computer Science Department
University of California at Santa Cruz
and
[§]Center for Evolution, Complexity and Cognition
Vrije Universiteit Brussel, Belgium
and
[†]Los Alamos National Laboratory
Research Library

http://www.soe.ucsc.edu/~okram
okram@soe.ucsc.edu

## Abstract

Pre-print repositories have seen a significant increase in use over the past fifteen years across multiple research domains. Researchers are beginning to develop applications capable of using these repositories to assist the scientific community above and beyond the pure dissemination of information. The contributions set forth by this paper emphasize a deconstructed publication model in which the peer-review process is mediated by an OAI-PMH peer-review service. This peer-review service uses a social-network algorithm to determine potential reviewers for a submitted manuscript and for weighting the influence of each participating reviewer's evaluations. This paper also suggests a set of peer-review specific metadata tags that can accompany a pre-print's existing metadata record. The combinations of these contributions provide a unique repository-centric peer-review model that fits within the widely deployed OAI-PMH framework.

## 1. Introduction

In 1991 the Los Alamos National Laboratory (LANL) announced the xxx.lanl.org pre-print digital-archive for the physics community [1]. This initiated the pre-print revolution that continues to allow scholarly publications to effectively reach the general public before the laborious task of official publication is completed. With pre-print repository technology there came a sophistication of library mechanisms not offered by simple World Wide Web publishing. The LANL archive provided a unique medium for searching, archiving, and allowing scientists to self-publish [2]. Then, with the introduction of the Open Archives Initiative [3] and the Institutional Repository movement, many universities and online publishers started their own repository systems to which, amongst other items, pre-prints can be submitted [4]. The OAI Protocol for Metadata Harvesting (OAI-PMH) was announced in 2001 as a protocol for accessing resource (including pre-print) metadata from any OAI-PMH compliant repository [5]. The OAI-PMH federates OAI repositories around the world and has allowed for a host of new repository services.

The contribution of this paper is a particular OAI service provider focused on the peer-reviewing of those pre-prints made available by OAI pre-print data providers. The proposed peer-review service can harvest pre-prints from OAI repositories, solicit potential referees to review the pre-prints, aggregate and weight referee evaluations, and finally supply the original repository with metrics concerning the pre-print's certification. This paper is outlined as follows. Section 2 describes the peer-review process in its current form. Related research in alternative peer-review models is discussed in Section 3. The whole of Section 4 contains the main contributions set forth by this paper. The introduction to Section 4 discusses the work-flow model of the OAI pre-print providers, the OAI peer-review service, and the peer-review referees. In Section 4.1, the paper presents a social-network algorithm that determines peer-reviewers in a manuscript-specific manner. Section 4.2 suggests a set of XML metadata tags specific to the peer-review process that, after review is complete, can be harvested by the original OAI pre-print provider. In Section 4.3, a prototype demonstration uses the described algorithm to generate referees for this paper. The paper concludes with a short discussion of the proposed model's benefits in Section 5.

The literature frequently makes a distinction between the terms e-print and pre-print [6]. E-print refers to the general category of any electronic manuscript. This paper will use the term pre-print throughout the text to refer to an electronic manuscript that has been submitted to a digital-library repository and is awaiting certification of it the author's claims.

## 2. The Standard Peer-Review Model

Peer-reviewed journals are currently the bedrock of scholarly communication. The main purpose of the peer-review process is to provide a certification of the written claims made by the authors. This certification process is shaped by the interaction of three main actors. The *authors*, as a result of their research, generate written reports of their work and then actively seek a public forum for their distribution. The *editors* and *publishers* provide such a forum in the form of a journal that contains a collection of works within a particular research domain. In order to ensure that the research is up to par with the standards of the publishing journal, the editors rely on experts within the field to review the claims made by the authors. It is ultimately up to these expert *referees* (a.k.a. reviewers) to accept or reject the work with respects to the integrity of the research community and the journal's quality standards [7]. By providing a peer-review model and a publication medium, the publishers have integrated the certification and distribution aspects of scholarly communication. Other forms of public distribution, like pre-print repositories, currently exist, but without any explicit certification the readers are responsible for gauging the credibility of the work they read and ultimately use in their research [8]. This is an undue burden to place on the community at large when, by only using the expertise of a few referees, a human-filter can effectively provide the community a quality assurance of published results. The necessity for certification has kept the current peer-review model relatively unchanged in the face of modern advances in communication.

Further entrenching the current peer-review model is the means by which the community determines the relevance of its researchers and journals to the scholarly endeavor. Thomson ISI journal impact factors are generally accepted as the standard for judging the influence of a scientist within the scientific community [9]. Simply stated, a journal's impact is determined by the number of articles within that journal that are referenced by articles in other journals [10]. When a scientist publishes in a high-impact journal, the impact rating of the scientist (sometimes called author impact) is taken to increase irrespective of whether or not his particular paper contributes to the impact rating of the publishing journal [11]. Because of the community's journal-centric approach to scholarly communication, the role of the journal, its editors, and its publisher have an overwhelming influence in shaping future research trends. Within the current 'publish (in high-impact journals) or perish' scientific model, it is difficult for new certification methodologies to gain a foothold—even in light of increased publication efficiency and potential end product quality [12]. The next section will discuss related research on alternative peer-review models before moving into this paper's proposed model and implementation.

## 3. Alternative Peer-Review Models

Proposals have been made to implement a scholarly process in which each stage of a manuscript's life is maintained by different groups using different technologies [13, 14]. The multi-stage model deconstructs scholarly communication into five stages—registration of a manuscript, certification of a manuscript, awareness of a manuscript, archiving of a manuscript, and finally rewarding an author for their manuscript. Of interest to this section in particular, and this paper in general, is the certification stage. The certification stage of the manuscript's life deals with validating the worthiness of a registered manuscript. It is important to note that the peer-review process is just one of many potential validation methods within the certification stage. Alternatives to the standard peer-review model have been proposed in the literature of multiple scientific domains. These alternative models embrace the importance of the peer-review process as a quality control mechanism, but unlike the standard peer-review model, they instead decouple the intimate relationship between journal editors and the peer-review process. The alternatives mentioned below are a collection of unique approaches to peer-review certification.

A relatively recent advancement called the interactive journal concept takes advantage of the Internet as a medium for open public discussion [15]. The interactive journal concept proposes a two-stage review process in which 'discussion papers' are first reviewed in an open forum allowing anyone in the community to contribute commentary. After a thorough discussion and author revisions, the discussion paper, now refined beyond the author's initial conception, can undergo the scrutiny of the more rigorous standard peer-review process. The interactive journal concept advances the importance of community feedback in reducing the labor required of the referees whom are time and time again burdened with the task of transforming deficient work into community ready material. The use of a larger pool of community members participating in the transformation of a pre-print into a formal publication can help reduce the workload of peer-reviewers and, more importantly, provide the author with a more diversified evaluation of their pre-print. On the other hand, a completely open model can overwhelm the author with superfluous review information. This paper provides a compromise between a small peer-review committee and a fully open process by restricting the influence of the community members in a pre-print specific manner.

The Berkeley Electronic Press, or bepress for short, has initiated an Internet-based publication medium in which manuscript refereeing and journal submissions are semi-independent processes [16]. Authors are not required to submit to a particular journal, but instead can submit to a family of journals. A bepress service called the Author & Reviewers' Bank incites past authors of a bepress journal to review present submissions. Once the pre-print has been evaluated by the referees, the evaluation is used to determine the appropriate journal for publishing the manuscript. Because bepress is home to a growing number of journals, the focus of reviewing is not *whether* to publish the paper, but more a question of *where* to publish the paper. Publication time is greatly reduced by the use of this mediating service. For instance, a pre-print doesn't need to be resubmitted to multiple rejecting journals of decreasing quality to find its appropriate public venue. This mechanism reduces the overall workload on the community because it uses only one set of referees to ultimately publish a paper. In a similar vein, this paper's proposed process maintains a similar review-once publish-anywhere model. In addition, this paper proposes an automated means by which pre-prints and reviewers are connected to one another.

Finally, Interjournal, a publication forum developed by the New England Complex Systems Institute, is described as a self-organizing refereed journal [17]. Interjournal combines the qualities of the two previous models in that it provides an automated public forum for commenting on submitted preprints and it also serves as a broad-topic journal with multiple publishable sub-categories. The review process is open to the community and the editor serves as the final mediator whom decides whether to accept or reject the paper with respects to the comments of the participating reviewers. In relevance to this paper is Interjournal's semi-open process where the comments and evaluations of the reviewers have varying degrees of influence depending on the domain of the paper. This context-sensitive influence measure, which determines the most appropriate referees for a pre-print, automates much of the editor's role in the peer-review process.

## 4. A Repository-Centric Peer-Review Model

The onset of pre-print repositories and the rapid dissemination of information brought on by self-publishing have introduced few 'friction' mechanisms which prevent the use of false information by the community. The ease of dissemination brought on by pre-print repositories may however bring about a more fluid method for peer-reviewing articles. This paper provides a peer-review model that separates the certification aspect of scholarly communication from the distribution aspect. In the proposed model, OAI repositories serve as the distribution medium, while on the other hand, an overlay OAI peer-review service provides the necessary certification mechanism.

The peer-review system proposed by this paper is an OAI service provider that makes use of OAI pre-print repositories to retrieve appropriate pre-prints to review. The general system architecture is outlined in Figure 1 and is coarsely explained as follows. The peer-review service can poll any number of OAI repositories for pre-prints deemed appropriate to review. It is up to the particular peer-review service implementation to decide which pre-prints to harvest. For example, some implementations may select pre-prints that include no Journal-Ref metadata, are within a certain ACM classification, are author requested, have a high usage or citation count, etc. (Figure 1-1). As will be shown, the peer-review service can then automatically determine and solicit the most appropriate reviewers for the pre-print (Figure 1-2). Next, the solicited reviewers are able to use a web-interface to submit their pre-print evaluations (Figure 1-3). The peer-review service then generates the appropriate peer-review metadata for the now peer-reviewed pre-print and stores it in its OAI peer-review data repository (Figure 1-4). Finally, the peer-review service's OAI repository exposes the metadata so that the original OAI pre-print repository can harvest the peer-review certification metadata to augment the pre-print's record metadata (Figure 1-5). Stages 2, 3, and 4 are discussed in the subsections to follow.

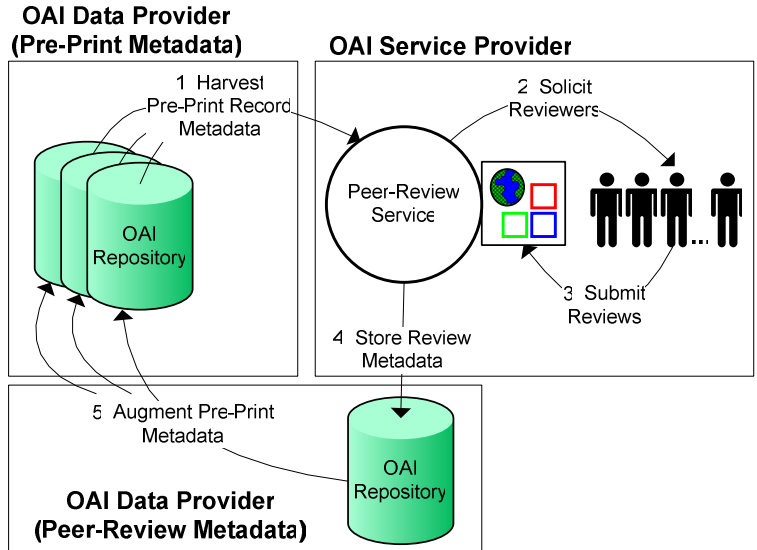

Figure 1. The architecture of the proposed repository-centric peer-review model

## 4.1 Soliciting and Weighting Peer-Reviewers

One of the major concerns of an editor is locating reviewers for a submitted manuscript. With an automated peer-review service, the standard pattern of locating three or four referees for a manuscript can easily be extended to incorporate a plethora of individuals competent in the various facets of a paper. For example, a paper such as this one pulls its references from multiple domains: social-network analysis, digital-library science, and scholarly communication models. Each submitted pre-print has a domain whether or not that domain is easily categorized within a single discipline or is intricately laid across many disciplines. This paper proposes that a pre-print's bibliographic references indicate its subject domains and the authors of those references are individuals who are competent in some aspect of that pre-print. Scanning a social-network for individuals related to the referenced authors may expose more individuals competent in the domain of the paper [18, 19]. A social-network can be created from many sources such as citation networks, co-citation networks, or an aggregate of multiple networks, but for the purposes of demonstration and proof of concept, a co-authorship network is used in this paper.

In a co-authorship network, a node represents an author and an edge between two authors represents a joint publication. In a repository-centric scholarly communication environment it is possible to construct co-authorship networks, for example, by harvesting the widely accepted Dublin Core `<dc:creator>` metadata tags from OAI repositories. A co-authorship network is defined by the data structure $G = [N, E, W]$, where $N$ is the set of nodes, $E$ the set of edges, $W$ the set of edge weights, and where $|E| = |W|$. For any one paper, the influence of that paper, $m_{i,j}$, on the edge weight between $author_i$ and $author_j$, $w_{i,j}$, is determined according to how many authors worked on that particular paper (Eq.(1): $m_{i,j} \in [0,1]$ ). For papers with many authors, the strength of tie between any two authors of that paper is weaker than a paper with few authors [18].

$$m_{i,j} = \frac{1}{x-1}$$

(1)

where $x$ is the number of authors of a particular paper

For all papers written by $author_i$ and $author_j$, $M$, the edge weight is the sum of all previously calculated co-authorship weights (Eq.(2): $w_{i,j} \in \mathbb{R}$ ).

$$w_{i,j} = \sum_{j=0}^{|M|} m_{i,j} \qquad (2)$$

After all the outgoing edge weights of a node have been determined, an edge probability is created by normalizing all outgoing values to 1.0 (Eq.(3): $w_{i,j_{(t)}} \in \mathbb{R}$ , $w_{i,j_{(t+1)}} \in [0,1]$ ).

$$w_{i,j_{(t+1)}} = \frac{w_{i,j_{(t)}}}{\sum_{x=0}^{|out(n_i)|} w_{i,x_{(t)}}} \qquad (3)$$

where out($n_i$) is the set of all outgoing edges to node $i$.

Once a weighted co-authorship network has been constructed, the next step is to match each harvested pre-print with an appropriate set of referees. In this proposed approach, this is a matter of determining the most appropriate reviewers given the bibliographic references of the pre-print. There are two ways to determine the bibliographic references of a pre-print. First, the OAI data provider may expose citation metadata. For instance, Biomedical Digital Libraries does just that [20]. Locating the resource of a record metadata has been explored at length in [21]. If such metadata does not exist then the actual pre-print object can be retrieved and parsed. Once the pre-print resource has been obtained, it can be analyzed automatically. For example, the Open Citation Project (OpCit) has developed a technology that is capable of extracting the bibliographic references from an e-print [22]. Supported e-print formats include HTML, PDF, PS, and plain-text. These extracted references are then parsed into their component parts (i.e. authors, journal, year, etc.).

Searching the co-authorship network for researchers related to the domain of the pre-print is implemented using a particle-swarm that begins at the referenced authors and then propagates throughout the co-authorship network—identifying the unique influence landscape of any submitted pre-print (Figure 3).

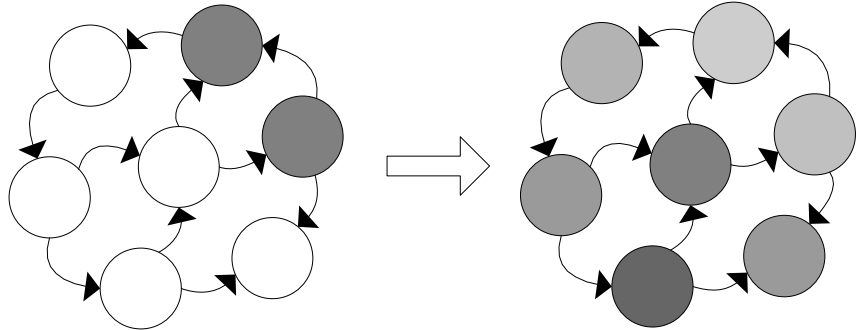

Figure 3. A particular subset of the network (authors referenced in pre-print) determines an influence distribution over the whole network (community)

The more a particular author is referenced, the more particles that author's node will initially receive. After the initial distribution, a stochastic particle-diffusion algorithm disseminates the particles throughout the co-authorship network allowing a particle to imprint its energy content on each node it traverses. Each particle, $p$, is endowed with an energy value, $\varepsilon$, where $p(\varepsilon) \in [0,1]$ . When a node receives a particle the node adds the particle's energy content to its memory, $n(\varepsilon)_{t+1} = n(\varepsilon)_t + p(\varepsilon)_t$ , where $n(\varepsilon) \in \mathbb{R}$ . A particle is a discrete indivisible entity, but a particle's energy content is a *not*. Therefore, each step that a particle traverses, from node to node, a certain portion, *ds*, of its energy content is decayed, $p(\varepsilon)_{t+1} = p(\varepsilon)_t - \left( p(\varepsilon)_t * ds \right)$, $ds \in [0,1]$. A high-level overview of the different stages of the particle-flow algorithm is presented in Table 1.

| 1: seed network | Each time an author in the reference section of the pre-print is mentioned, give his respective node in the co-authorship network 10 particles with energy content 1.0. |
| 2: propagate particles | Each particle of a node takes an outgoing edge determined by the probability distribution over all outgoing edges. |
| 3: add particle energy | Each node receiving a particle adds the energy content of the incoming particle to its current energy memory. |
| 4: decay particle energy | Each node decays the energy content of all its current particles. |
| 5: repeat steps 2-4 | Repeat steps 2-4 until all particle energy has decayed to 0.0. |

Table 1. A description of the stages of the particle-flow algorithm

Once every particle's energy content has decayed to 0.0, the algorithm is complete and the energy in each node represents the amount of 'influence' that the individual has relative to the pre-print. All node influence ratings are then normalized to derive each individual's proportion of influence relative to the whole collective (Eq (4): $n(\varepsilon)_t \in \mathbb{R}$, $n(\varepsilon)_{t+1} \in [0,1]$ ).

$$n_i(\varepsilon)_{t+1} = \frac{n_i(\varepsilon)_t}{\sum_{x=0}^{|N|} n_x(\varepsilon)_t} \qquad (4)$$

Each node in the network with any amount of influence can then be solicited to peer-review the pre-print. These influential nodes are the relatively-ranked experts with regards to the processed pre-print. The next subsection will describe how a referee's influence weights their evaluation relative to the collective evaluation. In this way, the particle diffusion algorithm serves both as a solicitation mechanism and a weighted influence mechanism.

Related research has proposed alternative metrics that utilize usage and co-authorship network analysis to determine an author's impact [23, 24]. What is unique about the proposed method's influence measure is that the influence of a referee is not a global measure but a value determined according to the domain of the manuscript and therefore well suited for the context-dependency seen in the peer-review process. The idea of domain specific influence is not new and has been called *local-trust* in social-network literature [25] and *relative-ranking* in literature on network influence metrics [26].

### 4.2 Peer-Review Metadata
When a pre-print is stored in an OAI repository it is associated with a metadata record. The URL below allows for the retrieval of this pre-print's record from the arXiv repository [27].

```
http://arxiv.org/oai2?
        verb=GetRecord&identifier=oai:arXiv.org:cs/0504084&metadataPrefix=oai_dc
```

The returned XML record contains metadata, `<metadata>`, formatted in the Dublin Core schema (`oai_dc`). For the sake of brevity, the full abstract of the paper which is located in the `<dc:description>` tag has been omitted.

```
<record>
  <header>
    <identifier>oai:arXiv.org:cs/0504084</identifier>
    <datestamp>2005-04-23</datestamp>
    <setSpec>cs</setSpec>
  </header>
  <metadata>
    <oai_dc:dc xmlns:oai_dc="http://www.openarchives.org/OAI/2.0/oai_dc/"
xmlns:dc="http://purl.org/dc/elements/1.1/" xmlns:xsi="http://www.w3.org/2001/XMLSchema-
```

```
instance" xsi:schemaLocation="http://www.openarchives.org/OAI/2.0/oai_dc/
http://www.openarchives.org/OAI/2.0/oai_dc.xsd">
      <dc:title>The Convergence of the Digital-Libraries and the Peer-Review
Process</dc:title>
      <dc:creator>Rodriguez, Marko A.</dc:creator>
      <dc:subject>Digital Libraries</dc:subject>
      <dc:subject>Computers and Society</dc:subject>
      <dc:description>Digital-libraries have seen a significant increase in use over the
past fifteen years across multiple research domains.
      </dc:description>
      <dc:date>2005-04-18</dc:date>
      <dc:date>2005-04-22</dc:date>
      <dc:type>text</dc:type>
      <dc:identifier>http://arxiv.org/abs/cs/0504084</dc:identifier>
      </oai_dc:dc>
   </metadata>
</record>
```

This paper proposes the inclusion of a set of peer-review XML tags to be used in parallel with other record metadata. The three main tags associated with the proposed peer-review schema (`pr`) are `pr:review`, `pr:referee`, and `pr:comment`. It is the role of a peer-review service and the solicited referees to support the creation of these tags. The `pr:referee` tag is a child tag of `pr:review` and represents a referee's identity and comments on a pre-print. The `pr:referee` tag contains three attributes which identify the referee's name, his influence in the peer-review process relative to the particular pre-print in review, and his current evaluation of the pre-print's quality.

```
<pr:referee name="Heylighen, Francis" influence="0.076" evaluation="0.65" />
```

If the community using the particular instantiation of the peer-review service desires a single-blind process, server-side processing can provide a persistent 'anonymous' value for the name attribute. The influence attribute is a pre-print specific value that is assigned according to the algorithm previously presented in Section 4-1. The evaluation property is the score that the referee has attributed to the pre-print. This value can range between 0.0 and 1.0 and serves as the referees subjective evaluation of the quality of the pre-print. Furthermore, the evaluation score can change throughout the course of the review process as the referee's comments affect the author's future version submissions. Notice that there is no accept/reject demarcation. The rational for this is that once a pre-print has made itself into a repository it has already been 'accepted' by the community. It is up to the community to gauge the relative influence and degree of visibility this pre-print should have through time. If a pre-print wishes to make its way to a more standard journal then it would be up to the journal to determine what score is acceptable according to the journal's quality standards. In this way, all pre-prints are available to the community, and therefore this model removes the sense of censorship perceived in the standard journal peer-review model [28].

Nested within the `pr:referee` tag is any number of `pr:comment` tags which specify comments that the referee wishes to share with the author of the pre-print. Comments are the only freestyle interaction that a referee can have with an author.

```
<pr:referee name="Heylighen, Francis" influence="0.076" evaluation="0.65" />
   <pr:comment data="2005-11-30">
   Your description of the 'particle-swarm' algorithm is not well explained. Your math
formalisms are not clear and the overall subsection is poorly organized.
   </pr:comment>
</pr:referee>
```

The current collective-subjective evaluation of a paper is identified by the `pr:review` tag which serves as the container tag for all peer-review specific data. The `pr:review` tag's attributes are evaluation and stability. The evaluation property refers to the collective's realization of the pre-print's quality and the stability property refers to the likelihood of the evaluation score to fluctuate given more reviewers.

By knowing the influence of a referee relative to the submitted pre-print, $inf$(n), and the reviewer's subjective evaluation, $eval$(n), the `pr:review` evaluation attribute value, $E$, for the paper can be determined as the weighted average of all actively participating referee, $A$, reviewing scores (Eq. (5):

$E \in [0,1]$; $inf(n) \in [0,1]$; $eval(n) \in [0,1]$, $|A| \leq |N|$). A weighted average of reviewer scores may not be the only possible method for determining the evaluation of a pre-print as extreme review scores may need to be curtailed. For simplicities sake, only the following weighted average is provided.

$$E = \frac{\sum_{x=0}^{|A|} inf(n_x) \bullet eval(n_x)}{\sum_{x=0}^{|A|} inf(n_x)} \tag{5}$$

Let's assume that a single less influential referee has scored a pre-print highly. At that point, the pre-print's evaluation is equal to that single referee's score. This is not a good measure of the pre-print's potential or future evaluation since a highly influential member can come into the review at some later time and dramatically alter the evaluation. For this reason it is important to add a stability measure to the review of the pre-print. The stability measure identifies the inverse of the likelihood that the pre-print's evaluation will change significantly given more referee evaluations. To put it another way, the stability value signifies how much of the total referee energy in the network has been associated with an evaluation. The pre-print's stability, *S*, is determined as the proportion of influence that the set of actively participating reviewers, *A*, have relative to the entire referee collective (Eq.(6): $S \in [0,1]$)

$$S = \sum_{x=0}^{|A|} inf(n_x) \tag{6}$$

If every individual in the community with some fraction of influence regarding the pre-print has participated in the review process then the stability of the pre-print is 1.0 and therefore stable. Stability is an important score because it provides a gauge for how many experts in the domain have reviewed and evaluated the pre-print and therefore, if the pre-print is deemed ready for use by the community.

The original OAI pre-print data provider can poll all its known OAI peer-review data providers for peer-review metadata associated with its stored pre-prints. For example, the following metadata request on the peer-review service's OAI repository returns peer-review certification metadata.

```
http://peer.review.service.org/oai2?
        verb=GetRecord&identifier=oai:arXiv.org:cs/0504084&metadataPrefix=pr

<record>
   <header>
      <identifier>oai:arXiv.org:cs/0504084</identifier>
      <datestamp>2005-04-24</datestamp>
      <setSpec>cs</setSpec>
   </header>
   <metadata>
      <pr:review evaluation="0.755" stability="0.50">
         <pr:referee name="Heylighen, Francis" influence="0.076" evaluation="0.65" />
            <pr:comment data="2005-11-30">
               Your description of the `particle-swarm' algorithm is not well explained.  Your
math formalisms are not clear and the overall subsection is poorly organized.
            </pr:comment>
         </pr:referee>
         <pr:referee>
         …
         </pr:referee>
      </pr:review>
   </metadata>
</record>
```

This metadata can then be used to augment the original OAI pre-print data provider's metadata for the pre-print. It may be of interest for the original repository to store, in parallel with other metadata, the entire `<pr:review>` section which includes referees and their comments, just the `<pr:review>` and `<pr:referee>` tags which omits the comments, or only the `<pr:review>` tag itself which includes only the collective evaluation and stability score.

### 4.3 A Prototype Demonstration

Before presenting the prototype's referee recommendations for this paper it is important to take a moment to discuss the history of this paper. The original idea and draft of this paper was initially written by Marko A. Rodriguez in Brussels, Belgium and was submitted to the Journal of Information Science under a single authorship. The journal editor, Adrian Dale, used Herbert Van de Sompel, the highest ranked reviewer provided by the prototype (as published in the pre-print), as a peer-reviewer of the manuscript (Table 2). Interestingly enough, due in part to the work described here, Marko left Belgium during the review process to begin a position at the Research Library at Los Alamos National Laboratory where both Herbert and Johan Bollen currently work. A peculiar situation was present where the peer-reviewer had so much to offer to the paper that a collaboration and ultimately a co-authorship emerged. We believe that this situation provides the first real-world validation of this algorithm.

With respects to this paper and its context in the scientific community, a successful scientist studying parallel computing should have little to no influence relative to some less accomplished computer scientist studying digital-library technology. The particle-swarm algorithm was run on the DBLP (Digital Bibliography and Library Project) [29] co-authorship network as of April 2005 and the results are displayed in Table 2. The potential referees are ordered according to their influence rating and their domain of interest is also supplied. The domain of interest was determined manually by analyzing each referees publications in the DBLP bibliography repository and locating those papers which are relevant to this work.

| Referee Name | Influence | Recent Interests Related to Paper |
|---|---|---|
| Sompel, HV | 0.09844 | OAI-PMH and Co-Authorship Networks |
| Bollen, J. | 0.08594 | Digital-Libraries and Network-Based Impact Metrics |
| Carr, L. | 0.08516 | Digital-Libraries and Open Archive Services |
| Hall, W. | 0.08066 | Knowledge Management and Digital-Libraries |
| Rocha, L.M. | 0.07892 | Document Recommendation Systems |
| Lagoze, C. | 0.05328 | OAI-PMH and Digital-Library Architectures |
| Harnad, S. | 0.04883 | Open Citation Linking and Digital-Library Architectures |
| Hitchcock, S. | 0.04177 | Electronic Journals and Citation Linking |
| Blake, M. | 0.04156 | OAI Repositories and Citation Linking |
| Jiao, Z. | 0.03386 | E-Print Services |
| Bergmark, D. | 0.03262 | Digital-Libraries and OAI-PMH |
| Miles-Board, T. | 0.02049 | Digital-Libraries |
| Davis, H.C. | 0.01211 | Digital-Libraries and Adaptive Linking |
| Roure, D.D. | 0.01125 | Dissemination of Scientific Information Services |
| French, J.C. | 0.01081 | Digital-Library Distributed Searching and Interfaces |
| Bailey, C. | 0.01043 | Digital-Libraries and Distributed Media |
| Brody, T. | 0.00986 | OAI-PMH and Open Citation Linking |
| Hunter, J. | 0.00950 | XML Schemas for Metadata Applications |
| Dushay, N. | 0.00911 | Metadata and Federated Digital-Libraries |
| Grange, S. | 0.00762 | Digital-Libraries |
| Millard, D.E. | 0.00737 | Digital Documents and Navigation |
| Fielding, D. | 0.00692 | Digital-Library Services |
| Payette, S. | 0.00648 | Metadata Architectures for Digital-Libraries |
| Hey, J. | 0.00571 | Open Citation Linking and Digital-Libraries |
| Woukeu, A. | 0.00477 | Dynamic Review Journals |

Table 2. the top 25 most influential referees with regards to this paper

Looking at Table 2, it is apparent that the algorithm should not allow authors to become peer-reviewers. A simple unimplemented, and therefore untested, solution is to provide the node's of the authors of the manuscript with 'negative' energy particles, $p(\varepsilon) \in [-1,0]$. This reduces the influence of the authors and those scientists for which those authors have co-authored with. Such a mechanism may combat potential 'conflict of interest' situations.

Qualitative methods are necessary to evaluate the effectiveness of reputation algorithms because the caliber/value/usefulness of the results is inherently subjective. For instance, the legitimacy of Google's PageRank algorithm [30] has been validated by its domination of the Internet search engine market. No explicit metric states that PageRank is the most optimal way to determine web-page reputation. Therefore, this paper will advance that this is not the only social-network metric that is viable for this particular application. Other algorithms do exist that could potentially serve as another means for determining referee influence. For example, the algorithm for detecting author similarity that has been presented in [31] is one such alternative method.

Future work in this area will focus on the large-scale validation of this algorithm. Currently, our research group is moving towards securing peer-review data from a particular publisher. This publisher's peer-review model allows referees to select which manuscripts they wish to review. The validation will be to determine the strength of correlation between those manuscripts selected by reviewers and those reviewers recommended by our algorithm. A future publication will contain this information.

## 5. Conclusion

The journal editor has had a long standing role as the primary mediator of the peer-review process. Accepting manuscript submissions, finding referees, aggregating reviews, and finally accepting or rejecting a manuscript have been implemented mainly as a manual human-driven process. With a combination of the peer-review system proposed by this paper and other OAI technologies, much of the editor's role in the peer-review process can be automated and further extended to meet the needs of the scientific community. A full-scale implementation of this system can provide the author the benefits of a semi-open community review-process, the referees the benefits of a review-once publish-anywhere environment, and finally, greatly reduce the responsibilities of the journal editor.

An important issue to discuss is the concept of 'incentive' as it applies to the peer-review process. Currently there exists no tangible incentive for referees to review manuscripts. The peer-review process is, at minimum, a single-blinded process and therefore the general-public is not privy to the names and evaluations of a manuscript's reviewers. The current peer-review model survives because scientists believe that peer-review is a responsibility that each and every member of the community must partake in if the scholarly process is to endure. With a computer-mediated peer-review service, the contributions of the peer-reviewers can actually be quantitatively represented and therefore used as another metric for gauging the relative influence that a scientist has in the community. A scientist may not be so well published, but instead may contribute a great deal to the community in their ability to peer-review literature. From here, it may be possible to derive other measures of scientific influence beyond the standard ISI impact factors. Reviewer metrics, such as the one proposed in [32] can be incorporated to create yet another layer of certification: reviewer certification.

## 6. Acknowledgements

Thank you very much to the referees whom reviewed this paper and, in doing so, made it significantly better than its initial conception. If that statement appears far-fetched, take the time to review this paper's pre-print rendition: http://arxiv.org/abs/cs/0504084v1. Also we would like to thank editors of the Journal of Memetics for adopting the prototype referee locator implementation for their review process [33]. Thank you to Francis Heylighen and Carlos Gershenson for extended talks on this topic. This research was supported by a GAANN Fellowship from the U.S. Department of Education, the Fonds voor Wetenschappelijk Onderzoek – Vlaanderen, and the Los Alamos National Laboratory.